\documentclass{article}
\usepackage{spconf,amsmath,graphicx}
\usepackage{subfigure}
\usepackage{algorithm}
\usepackage{algorithmicx}
\usepackage{algpseudocode}
\usepackage{booktabs} 
\usepackage{amssymb}

\floatname{algorithm}{Algorithm}

\title{GraphPB: Graphical Representations of Prosody Boundary in Speech Synthesis}
\name{Aolan Sun, Jianzong Wang\textsuperscript{*}\thanks{*Corresponding author: Jianzong Wang, jzwang@188.com}, Ning Cheng, Huayi Peng, Zhen Zeng, Lingwei Kong, Jing Xiao}
\address{Ping An Technology (Shenzhen) Co., Ltd.}
\begin{document}
%
\maketitle
\begin{abstract}
This paper introduces a graphical representation approach of prosody boundary (GraphPB) in the task of Chinese speech synthesis, intending to parse the semantic and syntactic relationship of input sequences in a graphical domain for improving the prosody performance. The nodes of the graph embedding are formed by prosodic words, and the edges are formed by the other prosodic boundaries, namely prosodic phrase boundary (PPH) and intonation phrase boundary (IPH). Different Graph Neural Networks (GNN) like Gated Graph Neural Network (GGNN) and Graph Long Short-term Memory (G-LSTM) are utilised as graph encoders to exploit the graphical prosody boundary information. Graph-to-sequence model is proposed and formed by a graph encoder and an attentional decoder. Two techniques are proposed to embed sequential information into the graph-to-sequence text-to-speech model. The experimental results show that this proposed approach can encode the phonetic and prosody rhythm of an utterance. The mean opinion score (MOS) of these GNN models shows comparative results with the state-of-the-art sequence-to-sequence models with better performance in the aspect of prosody. This provides an alternative approach for prosody modelling in end-to-end speech synthesis. \\
\end{abstract}
\section{Introduction}
\label{sec:intro}

In the field of neural text-to-speech, prosody is a crucial factor to determine intelligibility and naturalness of synthesised speech. Prosody can be refined as three suprasegmental features, fundamental frequency, loudness, and duration \cite{carmichael2002modeling,Aubin2019}. In the task of prosody modelling in neural text-to-speech, \cite{wang2017uncovering, skerryryan2018towards} firstly try to introduce a latent vector for prosody embedding which is extracted from the mel-spectrograms of the reference audios, then a global style token is introduced to be trained by multi-head attention for style encoding \cite{wang2018style,DBLP:conf/slt/StantonWS18}. Variational Auto Encoder (VAE) \cite{8683623} is tried to be used for prosody classifying for good prosody control. \cite{8683682} learns a latent embedding space of emotion derived from a desired emotional identity in a multi-speaker system. In order to achieve more precise local prosody control, \cite{lee2019robust} introduces temporal structure to enable fine-grained control of the speaking style of the synthesised speech. \cite{Koriyama2019} tries a semi-supervised speech synthesis framework in which prosodic labels of training data are partially annotated. \cite{bian2019multi} introduces a novel multi-reference structure to Tacotron to extract and separate different classes of speech styles: speaker, emotion and prosody.

Most of these methods try to analyse the prosody of a sentence from the speech-side, specifically exploiting prosody embedding from spectrograms of audios\cite{DBLP:conf/slt/LiuYWL18}. With the development of the pre-trained models like Bidirectional Encoder Representation from Transformers (BERT) \cite{devlin-etal-2019-bert} in the Natural Language Processing (NLP) tasks, some research is conducted on analysing prosody from text-side of text-to-speech. \cite{Hayashi2019, Yang2019} try to use BERT to encode input phrases, as an additional input to a Tacotron2-based sequence-to-sequence TTS model. However, the connection relationship between words is not fully reflected, which is an important factor that affects the rhythm and emotion of synthesised speech. \cite{Guo2019} firstly proposes to exploit the information embedded in a syntactically parse tree information to further improve the TTS quality. \cite{DBLP:conf/icassp/SunWCPZX20} proposes GraphTTS to model character-level and phoneme-level graph structure of the input utterance .The prosody boundary information \cite{chu-qian-2001-locating} is tried to be embedded into the Tacotron \cite{Wang2017, shen2018natural} model by \cite{8682368} through adding a context encoder analysing the context information. But the two separate Tacotron-like encoders project context features into two different domains. This may result in the exposure bias during model training and inference.
\begin{figure}[t]
    \centering
    \includegraphics[scale=3.15]{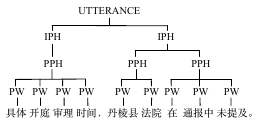}
    \caption{Prosody boundary}
    \label{fig:PB}
\end{figure}

Chinese utterance can be analysed in a hierarchical structure shown in Figure \ref{fig:PB}, which is defined according to the pause duration among phrases of a sentence. This can be manually labelled by human annotators, which is also open-sourced in some Chinese corpus. Four levels of prosody boundaries can also be seen a type of classification problem that different levels of boundaries can be predicted by the prosody boundary prediction model, which is a hot topic recently in the field of Chinese text-to-speech\cite{8682770, Pan2019}. Graph neural network can make use of the graph structure of prosody boundaries to hierarchically structure prosodic information. Potential models of the graph encoder in the task of text-to-speech can be graph convolutional network (GCN), gated graph neural network (GGNN), graph long-short term network (GLSTM), etc. This paper utilises GNN to model the hierarchical structure of Chinese prosody boundaries for context modelling. The contributions of this paper are:
\begin{itemize}
    \item Proposes two solutions to construct graph embedding by different levels of hierarchical prosody boundaries;
    \item Provides two approaches of phonetic sequential modelling, sequential edges and graph sequential encoder;
    \item Provides an alternative framework for text-to-speech, namely text-to-graph and then graph-to-speech modelling, to incorporate the prosody modelling module into the end-to-end speech synthesis process.
\end{itemize}

\section{Related Work}
Text-to-speech, a procedure to make talking machines, has been a developing hot topic in recent years because of the development of deep learning methods. Chinese is an ideogram language that each character has no relation with its phoneme, which is different from the phonetic language like English. However, the interrelation among words significantly influences the speaking rhythm of an input utterance.

For the purpose of visualisation of prosody embeddings,  \cite{Tits2019} tries to visualise and interpret these latent expressive variables through clustering plots. However, context features of texts are also essential to prosody modelling. The Recurrent Neural Network (RNN) encoder extracts parts of context information. \cite{8683857} has conducted experiments on investigating the similarity and difference between encoder outputs of the end-to-end system and the context information of the statistical parametric TTS. The experiment results show that the encoder outputs reflect both linguistic and phonetic contexts such as vowel reduction at phoneme level, lexical stress at syllable level, and part-of-speech at word level.  Prosodic words are the basic unit of a sentence. The consecutive words of the sentence form a prosodic word. Prosodic phrases are mostly composed of 2 or 3 prosodic words. Intonation phrases are separated by punctuation marks, such as commas, semicolons, etc. 

Graph Neural Network (GNN) is an effective solution for modelling complex relationships and interdependency between objects \cite{9046288, song2018graph}. Gated recurrent unit and Long-short term memory models are two effective approaches in the field of sequential modelling methods. The design of the forget gate is the essence of these two model. Similar gates are added in the graph neural network models in GLSTM and GRU. Li et al. 
The inputs are firstly converted into a graph and then the GNN are utilised to learn the representation from the input graph. The gated graph neural network (GGNN) uses the Gate Recurrent Units (GRU) in the propagation step, which is designed for sequential problems \cite{li2016gated}. \cite{zhang2018sentence-state, liang2016semantic} proposed Graph Long Short-Term Memory network (G-LSTM) to address the text encoding and semantic object parsing task. They follow the same idea of generalising the existing LSTMs into the graph-structured data in the non-Euclidean domain. The graph-to-sequence models have already been tested in the field of Neural Machine Translation (NMT) and Abstract Mean Representation (AMR) which shows outperforming the sequence-to-sequence models\cite{beck2018graph-to-sequence,  bastings2017graph, kong2017dragnn}. 

\subsection{Graph convolutional network}
Graph convolutional networks (GCNs) aim to generalize convolutions to graph domain. The methods in this direction are often categorized as spectral approaches and spatial approaches. GCN is a procedure to aggregate information from neighbourhood via a normalized Laplacian matrix. The shared parameters are from feature transformation. 

In the task of text-to-speech, since the input features are an undirected graph denotes by $X$, the nodes neighbouring to it can be denoted by the adjacency matrix $A$. The propagation procedure can be denoted by the equation below.

$$f(X, A) = A Relu(AXW^{(0)})W^{(1)}$$


This convolutional process can be regarded as a feature extraction process. In the field of speech synthesis, features are extracted from the input features which is used to be calculated attention matrix with the decoding spectrograms. However, the sequential relationships cannot be reasonably modelled through GCN, which might result in weak performance on long-term sentence.

\subsection{Graph recurrent network}

The GGNN model is designed for problems defined on graphs which require outputting sequences which is quite suitable for the speech synthesis task\cite{beck2018graph-to-sequence}. LSTMs are also used similarly as GRU through the propagation process based on a tree or a graph. Two types of Tree-LSTMs and graph-structured LSTM are proposed to address different tasks. They all follow the same idea of generalizing the existing LSTMs into the graph-structured data but has a specific updating sequence. A sentence-LSTM is also proposed for improving text encoding. It converts text into a graph and utilizes the Graph-LSTM to learn the representation.




\section{Graph Embedding}
The graph embeddings are defined according to a graph structure $  \mathcal{G} = (\mathcal{V}, \mathcal{E}) $, where $ v \in \mathcal{V}$, take unique values from $1, ..., \left| \mathcal{V} \right|$, and edges are pairs $e =(v, v^\prime) $. we focus on undirected edges in this paper represented by the directed edge $ v \to v^\prime $ and the reverse one $ v^\prime \to v $. The \textit{node embedding} (or \textit{node representation} or \textit{node vector}) for node $v$ is denoted by $\boldsymbol{h}_v \in \mathbb{R}^D$ \cite{scarselli2008graph, zhou2018graph}.

\subsection{Prosody boundary modelling}
\label{sec:pbge}

\begin{figure}[h]
    \centering
    \subfigure[$PPH_1$]{%
      \includegraphics[scale=0.93]{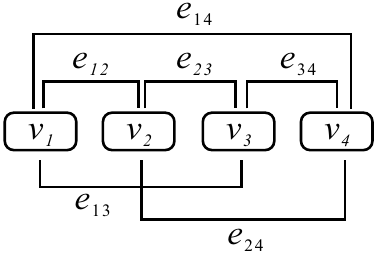}
      \label{fig:subfigure1}}
    \quad
    \subfigure[$PPH_2$]{%
      \includegraphics[scale=0.93]{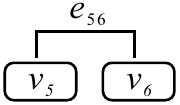}
      \label{fig:subfigure3}}
    \quad
    \subfigure[$PPH_3$]{%
      \includegraphics[scale=0.93]{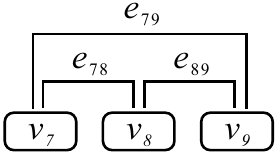}
      \label{fig:subfigure2}}
    \quad
    \subfigure[$IPH_2$]{%
      \includegraphics[scale=0.93]{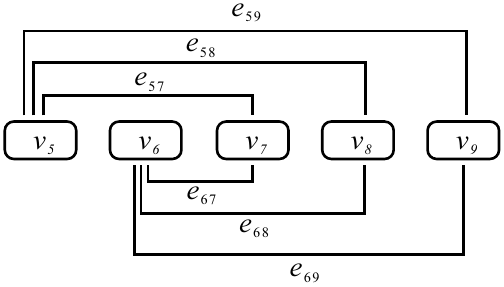}
      \label{fig:subfigure4}}
    \caption{Graphical prosodic boundary graph embeddings}
    \label{fig:GPB}
\end{figure}

The graph embeddings are constructed according to the hierarchical prosody boundary structure, prosodic word (PW), prosodic phrase  (PPH), intonation prosodic phrase (IPH) and utterance (utter) denoted by $\mathcal{L}_{pw}, \mathcal{L}_{pph}, \mathcal{L}_{iph}, \mathcal{L}_{utter}$. There is a containment relationship that shows in the equation below.
\begin{equation}
    \mathcal{L}_{pw} \subseteq \mathcal{L}_{pph} \subseteq \mathcal{L}_{iph} \subseteq \mathcal{L}_{utter}
\end{equation}{}
The nodes are composed of PWs that $v = \mathcal{L}_{pw}$ because the PWs are the minimum units of the prosody boundaries. The relationship of the PWs, PPHs and IPHs is heterogeneous, which makes it difficult to model edges among different level's nodes. So we model the containment relationship among different levels of prosody boundaries as edges $e \in \mathcal{E}$. 
\begin{equation}
    v = \mathcal{L}_{pw} \in \mathcal{V}
\label{eq:2}
\end{equation}{}
\begin{equation}
    e = \left\{ \mathcal{L}_{pph}, \mathcal{L}_{iph}, \mathcal{L}_{utter}\right\} \in \mathcal{E}
\label{eq:3}
\end{equation}{}

The prosody boundaries of the Figure \ref{fig:PB} is converted to graph-structured shown in the Figure \ref{fig:GPB}. To be specific, because the second PPH in the Figure \ref{fig:PB} composed of the $5^{th}$ and $6^{th}$ PWs, there is an edge connecting nodes $v_5$ and $v_6$, denoted by $e_{56} = (v_5, v_6)$ shown in Figure \ref{fig:subfigure3}. Similarly, the graph embeddings of $PPH_1$ and $PPH_3$ constructed are shown in Figure \ref{fig:subfigure1} and Figure \ref{fig:subfigure2}. In Figure \ref{fig:PB}, the second IPH comprises of $PPH_2$ and $PPH_3$. So the $IPH_2$ can be modelled as in Figure \ref{fig:subfigure4} that six edges are connected from the PWs of $PPH_2$ to the PWs of $PPH_3$. Only one edge remains when pph-edge and iph-edge are duplicate.

\subsection{Sequential modelling}
\label{subsec:seqmod}
The sequential relationship among phonemes cannot be ignored because the significant variants of Chinese pronunciation like soft tone, transposed tone is highly related to the words connection scenarios. So it is necessary to add sequential representations into the graph embedding.

Two methods are proposed and experimented. One approach is to solve the problem in the process of making the graph embedding, that sequential edges, $e = (v_i, v_{i+1})$ where $i=1,...,n $, are added between adjacent nodes for better representation of the timing relationship. This may result in duplicate edges with the prosodic boundaries but this can be interpreted as a second pass of the messages in this scenario that the adjacent nodes may have a strong relationship than the not adjacent ones. An alternative approach is to add an additive encoder, which will be detailly described in Section \ref{subsubsec:g2s struc}.


\subsection{Algorithm}
The algorithm of constructing graph embedding is shown in Algorithm 1. The input is the sentence with prosody boundaries $S$ and the set of prosody boundaries $\mathcal{P}=\{P_1, P_2\}$. These prosody boundaries can be predicted by the prediction models or annotated manually. The output of the algorithm is graph embedding of the sentence $  \mathcal{G} = (\mathcal{V}, \mathcal{E}) $, which is an intermediate embedding as the input of the graph-to-sequence TTS model. The notations are shown in the Table\ref{tab:10}.

\begin{algorithm}[h]
    \caption{Graph Embedding Construction}
    \begin{algorithmic}[1] 
    
        \Require Sentence with prosody boundaries $S$; The set of prosody boundaries $\mathcal{P}=\{P_1, P_2\}$
        \Ensure Graph Embedding of the sentence $  \mathcal{G} = (\mathcal{V}, \mathcal{E}) $
        \State Initialize $b,bb,ss, pb,\mathcal{V}, \mathcal{E}\gets[], idx\gets0$
        \For{each $s \in S$}
            \State $cur \gets [b;s]$
            \If{$cur \in \mathcal{P}$}
                \State $idx \gets idx + 1$
                \State $pb \gets [pb;cur]$
                \State $ss \gets [ss;idx;cur]$
                \State $\mathcal{V} \gets [\mathcal{V};cur]$
            \EndIf
            \State $b \gets s$
        \EndFor
        
        \For{each $n \in ss$} 
            \If{$n \neq P_2$}
                \State $bb \gets [bb;n]$
            \Else
                \State $nns \gets bb$\textit{ segmented by }$P_1$
                \State $ee \gets $\textit{the combinations of two random elements in }$nns$
                \State $\mathcal{E} \gets [\mathcal{E};ee]$
                \State $bb \gets ''$
            \EndIf 
        \EndFor
            
            \State $result \gets (\mathcal{V}, \mathcal{E})$ 
        
        \State \Return{$result$}
    \label{algorithm}
    \end{algorithmic}
\end{algorithm}

\begin{table}[h]
  \caption{Notation of Algorithm 1}
  \label{tab:10}
  \centering
  \begin{tabular}{ c@{}l  l }
    \toprule
    \multicolumn{2}{c}{\textbf{Notation}} & \multicolumn{1}{c}{\textbf{Meaning}} \\
    \midrule
        $\mathcal{S}$  &      &    The sentence with prosody boundaries    \\
        $\mathcal{P}$  &      &     The set of prosody boundaries    \\
        $P_1$          &      &     The prosody-word boundary           \\
        $P_2$          &      &     The phrase-prosody boundary       \\
        $\mathcal{G}$  &      &     The graph embedding of the sentence    \\
        $\mathcal{V}$          &      &      The vertices of the graph embedding          \\
        $\mathcal{E}$          &      &     The edges of the graph embedding         \\
        $b$           &      &    The previous character    \\
        $ss$          &      &    $\mathcal{S}$ with characters replaced by placeholders       \\
        $pb$         &      &    The array of prosody boundaries  $\mathcal{S}$   \\
        $s$          &      &     The current character in the sentence $\mathcal{S}$      \\
        $cur$          &      &     Current scanning word      \\
        $idx$          &      &     The placeholder of the character     \\
        $bb$          &      &    The consecutive characters            \\
        $n$          &      &   The current character in the   $ss$         \\
        $nns$          &      &  The nodes connecting by this     $P_2$      \\
        $ee$          &      &    The edges connected by  $P_2$       \\
    \bottomrule
  \end{tabular}
\end{table}

\section{Graph-to-sequence TTS}
\label{sec:g2stts}
The graph-to-sequence structure is an alternative approach of sequence-to-sequence models \cite{DBLP:conf/nips/SutskeverVL14} in the task of the end-to-end process, which has shown effectiveness in the field of neural machine translation (NMT). In the task of text-to-speech (TTS), graph embedding is the information converted from sequential text to a non-Euclidean space, which requires a graph encoder to compute and pass the messages embedded in nodes and edges of the input graph.

\begin{figure*}[t]
    \centering
    \subfigure[Graph-to-Sequence TTS]{
    \begin{minipage}[c]{.3\textwidth}
    \centerline{\includegraphics[scale=0.85]{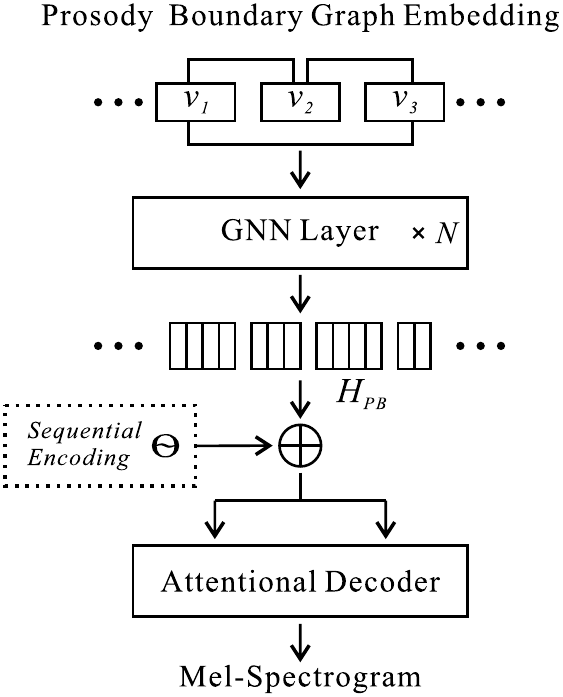}}
    \label{fig:g2stts1}
    \end{minipage}}
    \hfill
    \subfigure[Graph Sequential Encoder]{
    \begin{minipage}[c]{.3\textwidth}
    \centerline{\includegraphics[scale=0.85]{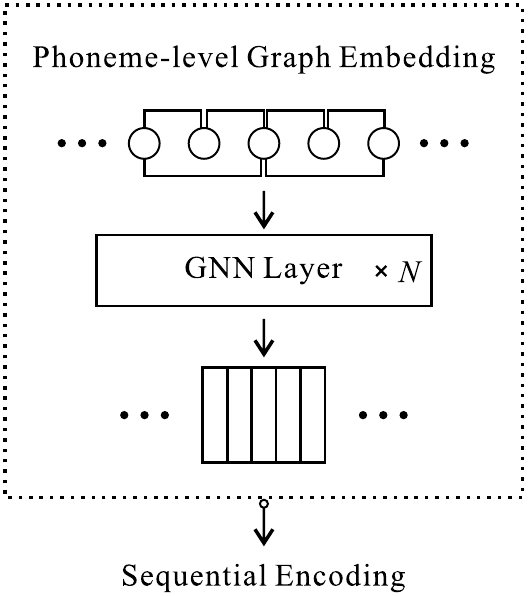}}
    \label{fig:seq_mode}
    \end{minipage}}
    \hfill
    \subfigure[Graph Neural Network Propagation]{
    \begin{minipage}[c]{.3\textwidth}
    \centerline{\includegraphics[scale=0.85]{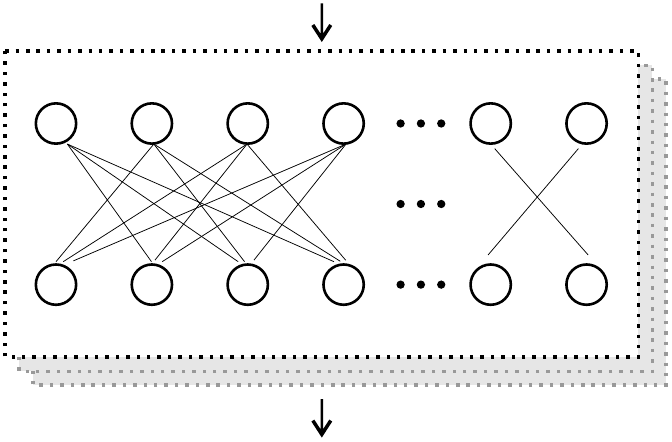}}
    \label{fig:gnn}
    \end{minipage}}
    \hfill
    \caption{Graph-to-sequence TTS}
    \label{fig:g2s TTS}
\end{figure*}


\subsection{Model Structure}
\label{subsubsec:g2s struc}
In the Graph-to-Sequence TTS model, the converted prosody boundary graph embeddings are first consumed by $N$ GNN layers to encode the embedding information. The propagation process of the model is shown in Figure \ref{fig:gnn}. The number of neurons of the first layer of GNN encoder is the same as the number of nodes of graph embeddings. The nodes embedding are randomly initialised and the information aggregates along the edges connected to it. The propagation process shown in Figure \ref{fig:gnn} is quite similar to the fully-connected neural network. However, the key difference is that the messages will only aggregate among adjacent nodes connected by graph edges. 

This propagation process will converge to a fixed point that nodes embedding achieves a steady state. In the task of speech synthesis, 2-3 layers of GNN is enough for text encoding. The Graph neural network (GNN) layer can make use of GGNN or G-LSTM models for this sequential task. The output of this GNN encoder is the value vector $H_{PB}$ attending the calculation process of the attentional decoding procedure same as the one used in Tacotron2. Sequential encoding calculated by the Graph Sequential Encoder is optionally concatenated with the encoder outputs $H_{PB}$ . Mel-spectrograms are output through the attentional decoder for generating waveforms. 

\subsection{Graph Sequential Encoder}
To solve the problem of sequential modelling, an additive encoder, \textit{Graph Sequential Encoder}, can be added for sequential modelling shown in Figure \ref{fig:seq_mode}. The input of the \textit{Graph Sequential Encoder} is the phoneme-level graph embeddings. The nodes of the phoneme-level graph embeddings are the phoneme characters of a sentence and the edges are the sequential edges connecting the characters representing the sequential information. These phoneme-level graph embeddings are input into $N$ GNN layers for encoding. Similar to the GGNN module in the graph encoder shown in \ref{subsubsec:g2s struc}, the Graph neural network (GNN) layer can make use of GGNN or G-LSTM for the prosody encoding task. The output sequential encoding $\Theta$ can be concatenated with the prosody boundary graphical representations $H_{PB}$ to attend the calculation process of attentional decoding as the keys and values of the attention mechanism. This may give the decoder more information about phoneme-level and sequential information, which is expected to improve the synthesis performance.

\section{Experiments}
\subsection{Training setup}
The experiments are conducted on the open-source dataset from the Databaker company \cite{Biaobei}. The dataset contains graphemes, phonemes, prosodic labels and the corresponding audios. The prosodic labels can also be predicted by the prosody boundary prediction models. The total hour of the dataset is about 12 hours with 10000 utterances recorded by a professional Chinese female news anchor. Because of the data sparsity of Chinese characters, the Chinese graphemes $\Psi$ are firstly converted to phonetic characters in a compact space denoted by $\psi$. The dimensions of $\psi$ are significantly smaller than that of $\Psi$.  The baseline model as the benchmark is Tacotron2. Griffin-Lim \cite{griffin1984signal} is selected as the vocoder for the mainly comparative experiment on prosody performance. The frame length and frameshift in the models are 50ms and 12.5ms, and the output acoustic features are 80-dim mel-spectrograms. Batch-size is $32$ and Adam optimizer is utilised. The learning rate follows a designed annealing strategy starting from $1e-3$ decreasing to $1e-5$ after $5000$ iterations. 

The subjective evaluation metric chosen in this paper is Mean Opinion Score (MOS), scaling from 0 - 5 with stages increased by 0.5. The listening tests are rated by 50 native speakers on 100 randomly chosen test sentences. Each sentence is scored by at least 10 raters. The MOS tests were crowdsourced to the raters through an internal platform similarly to Amazon's Mechanical Turk. 

\subsection{Experiment Design}
\subsubsection{Experiment I -- Benchmark experiment}
Experiment I is designed to show the technical feasibility of using GNN models to solve TTS problems using prosody boundary inputs. Two GNN approaches are selected as the encoder, GGNN and GLSTM respectively. Only PPH edges are modelled in this experiment for simplicity and sequential edges are added for avoiding missing edges in some utterances. The MOS results of the baseline model and GraphPB models as shown in the Table \ref{tab:1}.

\begin{table}[th]
  \caption{Benchmark experiment}
  \label{tab:1}
  \centering
  \begin{tabular}{ c@{}l  r }
    \toprule
    \multicolumn{2}{c}{\textbf{Model}} & 
                                         \multicolumn{1}{c}{\textbf{MOS}} \\
    \midrule
    Baseline Tacotron2               &      & $4.05 \pm 0.12 $~~~             \\
    GraphPB GGNN Encoder              &      & $4.28 \pm 0.22 $~~~               \\
    GraphPB GLSTM Encoder             &      & $4.25 \pm 0.26 $~~~       \\
    \midrule
    Human Recording               &      & $4.60 \pm 0.17 $~~~   \\
    \bottomrule
  \end{tabular}
\end{table}

From Table \ref{tab:1}, it can be seen that the GraphPB GGNN and GLSTM Encoder model achieve competitive performance with the baseline model. The model of GGNN Encoder achieves slightly higher MOS than GLSTM Encoder. However, the robustness of the two GraphPB models performs not that good as the baseline model. This may be due to the huge complexity of the edge information.

\begin{figure}[t]
    \centering
    \includegraphics[scale=0.45]{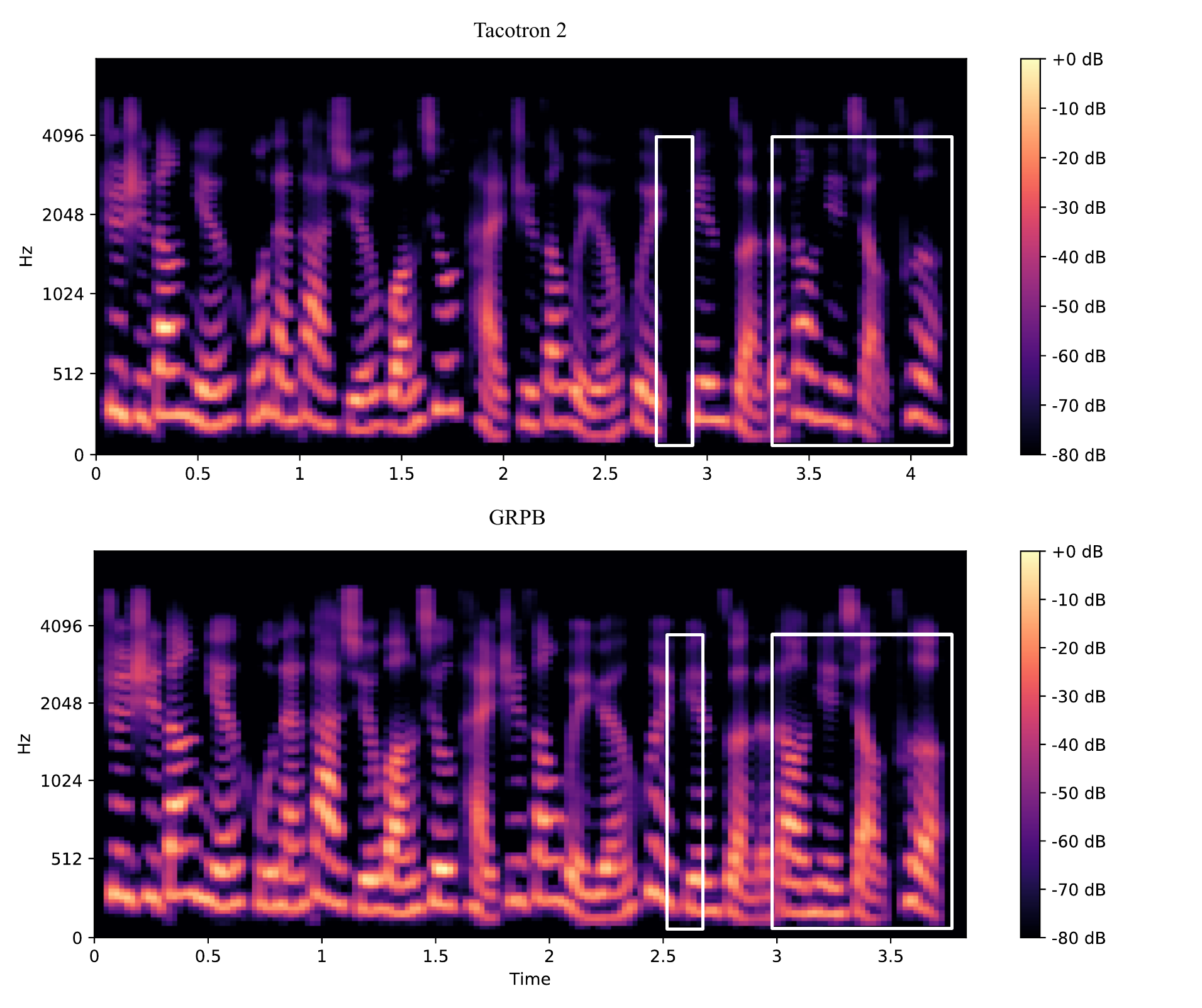}
    \caption{Comparison of Tacotron2 and  GraphPB}
    \label{fig:mels}
\end{figure}

The mel-spectrogram figures of the baseline Tacotron and GraphPB GGNN Encoder are shown in Figure \ref{fig:mels}. It can be seen that the pausing in Tacotron2 is more obvious, which is determined by the space symbols in the input text, whereas the pausing information in the GGNN Encoder model is more smooth among words, which gives a better flow of speaking rhythm. Besides the spectrogram of the end part of the GraphPB model has more energy than that of Tacotron2, which shows better rhythm at the end of the audio.

\subsubsection{Experiment II -- Sequential modelling}
Two approaches for encoding sequential information introduced in Section \ref{subsec:seqmod} and Section \ref{subsubsec:g2s struc} are compared in this experiment for the necessity of phoneme-level sequential modelling. The level of prosody boundary edges used in the experiment is PPH.

\begin{table}[th]
  \caption{Sequential modelling}
  \label{tab:2}
  \centering
  \begin{tabular}{ c@{}l  r }
    \toprule
    \multicolumn{2}{c}{\textbf{Model}} & 
                                         \multicolumn{1}{c}{\textbf{MOS}} \\
    \midrule
    No Sequential Info            &      & $3.97 \pm 0.35$~~~       \\
    Sequential Edges            &      & $4.28 \pm 0.22$~~~               \\
    Graph Sequential Encoder               &      & $4.30 \pm 0.19$~~~             \\
    \bottomrule
  \end{tabular}
\end{table}

It can be observed from Table \ref{tab:2} that the injection of sequential information improves the final results of the naturalness. The average MOS of graph embeddings without sequential info is the lowest with the highest variance. This may be due to high probabilities of null edges. The addition of Graph Sequential Encoder achieves the highest MOS and the lowest variance. But the model complexity is relatively high with low convergency efficiency.

\subsubsection{Experiment III -- Number of level of PBs}
Experiment III is designed to fine-tune the number of levels of prosody boundaries to be considered in graph embeddings. The encoder used in this experiment is GGNN Encoder and sequential edges are added. The experiment results are shown in Table \ref{tab:4}.

It is shown in Table \ref{tab:4} that the additive IPH information slightly improves the MOS results with a gap of 0.01 and the slight drop of variance. However, the duplicate edges and complex hierarchical structure results in low convergency efficiency.

\begin{table}[th]
  \caption{Number of levels of PBs}
  \label{tab:4}
  \centering
  \begin{tabular}{ c c  c }
    \toprule
    \multicolumn{1}{c}{\textbf{Prosody boundary}} & \multicolumn{1}{c}{\textbf{MOS}} & \multicolumn{1}{c}{\textbf{Convergency Steps}}\\
    \midrule
    PPH  &  $4.28 \pm 0.22$ &   $11k$~~~         \\
    PPH \& IPH  &  $4.29 \pm 0.21$ &   $23k$~~~            \\

    \bottomrule
  \end{tabular}
  
\end{table}

\subsubsection{Experiment IV -- Comparative analysis}
To deeply discover the representations of prosody boundary, one utterance is segmented through two different methods shown below. 


\begin{itemize}
    \item U1:  $\mathcal{V} = \{v_1, v_2\} ; \mathcal{E} = \{(v_1, v_2)\}$ 
    \item U2: $\mathcal{V} = \{v_1, v_2, v_3, v_4, v_5\} ; \mathcal{E} = \{(v_1, v_2), (v_4, v_5)\}$
\end{itemize}{}


U1 segments the utterance into two parts according to the comma and period punctuation, so that there are only two nodes $v_1, v_2$ representing each part and one edge connecting them, $(v_1, v_2)$. U2 segments the sentence to 5 parts represented by 5 nodes, $v_1, v_2, v_3, v_4, v_5$, and 2 PPHs are added in this version of segmentation represented by 2 edges, $(v_1, v_2), (v_4, v_5)$.

The level of prosody boundary to make graph embedding is PPH and the GGNN Encoder is used in this experiment. The MOS experiment results are shown in Table \ref{tab:5}. The MOS of U2 is higher with better robustness because more reasonable segmentation rules are applied according to the semantic rules of the utterance. The segmentation of U1 is in a low segmentation resolution which results in slightly lower MOS and higher variance. This empirically shows that the embedding of graphical representations of prosody boundary can improve the prosody performance of speech synthesis.

\begin{table}[th]
  \caption{Comparative analysis}
  \label{tab:5}
  \centering
  \begin{tabular}{ c@{}l  r }
    \toprule
    \multicolumn{2}{c}{\textbf{Utterance}} & 
                                         \multicolumn{1}{c}{\textbf{MOS}} \\
    \midrule
    U1               &      & $4.19 \pm 0.2$~~~             \\
    U2              &      & $4.25 \pm 0.12$~~~       \\
    \bottomrule
  \end{tabular}
  
\end{table}





\section{Conclusions}
This paper utilises Chinese prosody boundary to form graph embedding, that is consumed by a graph recurrent model for graph encoding. The encoded context features are passed to an attentional decoder for outputting mel-spectrogram frame-by-frame. The experiment results show competitive performance with the state-of-the-art sequence-to-sequence models in the spectrogram generation module. Similar graph construction techniques and graphical modelling approaches can be tested in other languages.

\section{Acknowledgements}
This paper is supported by the National Key Research and Development Program of China under Grant  No.2017YFB1401202, No.2018YFB0204400 and No.2018YFB1003500. The corresponding author is Jianzong Wang from Ping An Technology (Shenzhen) Co., Ltd.


\vfill\pagebreak
\bibliographystyle{IEEEbib}
\bibliography{mybib}

\end{document}